\documentclass[journal]{IEEEtran}

\newlength{\figwidth} 
\usepackage[english]{babel}
\usepackage{verbatim} 
\usepackage{varioref}
\usepackage{psfrag}
\usepackage{epsf}
\usepackage{graphics}
\usepackage{graphicx}
\usepackage{amsbsy}
\usepackage{amssymb}
\usepackage{amscd}
\usepackage{amsmath}
\usepackage{amsthm}
\usepackage{enumerate} 
\usepackage{color}
\usepackage{epstopdf}

\usepackage{cite}
\usepackage{url}

\def\editcolor{black} 

\title{Suburban Residential Building Penetration Loss\\ at 28 GHz for Fixed Wireless Access}

\author{ 
Jinfeng Du, \emph{Member, IEEE}, Dmitry Chizhik, \emph{Fellow, IEEE}, Rodolfo Feick, \emph{Senior Member, IEEE},  
Guillermo Castro, Mauricio Rodr\'iguez, \emph{Member, IEEE}, and Reinaldo A. Valenzuela, \emph{Fellow, IEEE}
\thanks{Jinfeng Du, Dmitry Chizhik and Reinaldo. A. Valenzuela are with Nokia Bell Labs, Holmdel, NJ 07733, USA (e-mail: 
\{jinfeng.du, dmitry.chizhik, reinaldo.valenzuela\}@nokia-bell-labs.com).}
\thanks{Rodolfo Feick is with Department of Electronics, Universidad T\'ecnica Federico Santa Maria, Valpara\'iso, Chile (e-mail: rodolfo.feick@usm.cl).}
\thanks{Guillermo Castro and Mauricio Rodr\'iguez are with Escuela de Ingenier\'ia El\'ectrica de la Pontificia Universidad Cat\'olica de Valpara\'iso, Valpara\'iso, Chile (e-mail: \{guillermo.castro, mauricio.rodriguez.g\}@pucv.cl).} 
\thanks{The authors wish to acknowledge support by CONICYT under Grant Proyecto Basal FB0821, Fondecyt Iniciaci\'on 11171159 and to Proyecto VRIEA-PUCV 039.462/2017 for supporting Mauricio Rodr\'iguez and Guillermo Castro.}
} 

\begin{document}

\maketitle

\begin{abstract}
 Fixed wireless access at mm/cm bands has been proposed for high-speed broadband access to suburban residential customers and building penetration loss is a key parameter. We report a measurement campaign at 28 GHz in a New Jersey suburban residential neighborhood for three representative single-family homes. A base antenna is mounted at 3-meter height, emulating a base station on a lamppost,  moves down the street up to 200 meters. A \textcolor{\editcolor}{customer premises equipment (CPE) device},  acting as relay to provide indoor coverage throughout the desired area, is mounted either on the exterior of a street-facing window or 1.5 meters behind the window.  The median indoor-outdoor path gain  difference, \textcolor{\editcolor}{corresponding to the extra loss by moving the window-mounted CPE indoor,} is found to be 9 dB for the house with low-loss materials and plain-glass windows, and 17 dB for the house with low-emissivity windows and foil-backed insulation. \textcolor{\editcolor}{These losses are in addition to  other  losses (e.g., vegetation blockage) in comparison to free space propagation.}
\end{abstract}
 


\section{Introduction}\label{sec:introduction}
 
The wide spectrum available at cm/mm bands and the ability to pack  more elements within the same antenna size promise very high rates for distance up to a few hundred meters. Fixed wireless access (FWA) at such bands, especially 28/39 GHz bands, has been proposed as an attractive solution~\cite{WCL_1} to provide high-speed broadband connection to suburban residential customers, especially for markets where the fiber-to-the-home is unavailable. However, the high rate promise of FWA {to indoor users} crucially depends on building penetration loss.

	Various measurement campaigns at mm/cm bands \cite{WCL_2, WCL_3, WCL_4, WCL_5, WCL_6, WCL_7, WCL_8, WCL_9, WCL_10} have been conducted in the past few years to assess penetration loss through different materials and buildings, and an extensive survey  can be found in \cite{WCL_10} and in \cite{WCL_9} (for frequencies below 6 GHz). For example, Zhao et al. \cite{WCL_4} reported  \textcolor{\editcolor}{more than 20 dB difference in} penetration loss through {plain and tinted glass} at 28 GHz. Larsson et al.  \cite{WCL_6} reported penetration loss at 28 GHz for an open office building with brick/concrete exterior wall and windows with standard/coated glass. The \textcolor{black}{penetration loss} at the measured locations varied from 3 dB to 60 dB, \textcolor{\editcolor}{which suggests that outdoor-to-indoor coverage in office buildings cannot be assured}. Haneda et al. \cite{WCL_8} surveyed building penetration loss, reporting that results are consistent with a penetration loss model based on composition of construction materials. The model was adopted by 3GPP \cite{WCL_11} where the low-loss model corresponds to a building with 70\% concrete and 30\% standard multi-pane glass. \textcolor{\editcolor}{Note that the 3GPP model specified in \cite{WCL_11} is aimed at facilitating simulation comparison, and it may not provide accurate prediction for all types of constructions.} For example,  in the United States wood and wood products account for more than 90\% market share for exterior/interior wall construction in single-family houses\cite{WCL_12}, resulting in far less penetration loss than concrete.
	Raghavan et al.~\cite{WCL_5} reported a median penetration loss of 9.2 dB for 22-43 GHz bands through strand board, \textcolor{\editcolor}{a modern construction material for exterior walls,  and commented that ``\emph{outdoor-to-indoor coverage is more likely in older/residential settings}''}.  
	Bas et al.~\cite{WCL_7} measured point-wise power difference between various outdoor and indoor locations at 28 GHz and reported an average loss of 10.6 dB for a \textcolor{\editcolor}{wood-frame} single-family home. \textcolor{\editcolor}{Power difference between contiguous locations inside and outside windows changes between 1 and 21.5 dB, which highlights  coverage uncertainty   at random indoor locations.}  
	
	\textcolor{\editcolor}{Given the observed challenges of outdoor-to-indoor coverage at 28 GHz \cite{WCL_4, WCL_5, WCL_6, WCL_7},} and the fact that FWA base stations are very likely to be mounted on lampposts along the streets~\cite{WCL_1}, customer premises equipment (CPE)  \textcolor{\editcolor}{is desirable} as relay to cover an indoor area. CPEs are more likely to be installed at some favorable location in the home than at  random indoor locations as measured by~\cite{WCL_6, WCL_7}. \textcolor{\editcolor}{Besides, the distance from CPE to base stations may vary from tens of meters to a couple of hundred meters, whereas~\cite{WCL_6, WCL_7} used a fixed distance.} Therefore, new measurements are necessary for such setups. 
	
	We report here a measurement campaign at 28 GHz in a New Jersey suburban residential neighborhood using a narrowband channel sounder. A fixed transmit horn antenna, mounted either on the exterior of a street-facing window to emulate a window-mounted CPE, or 1.5 meters behind the window (same height, mimicking a desktop modem). A spinning receiving horn antenna mounted at 3-meter height, emulating a base station on a lamppost, moves down the street up to a range of 200 meters.  Three representative single-family homes were measured: 1) conventional low-loss building materials with plain glass windows; 2) renovated building with low-emissivity windows; 3) new construction using foil-backed insulation and low-emissivity windows.

\section{Measurement setup and data processing}\label{sec:measurement}  

\subsection{Measurement equipment}\label{sec:setup} 

  To maximize link budget and data collection speed, we constructed a narrowband sounder, transmitting a 28 GHz \textcolor{\editcolor}{continuous-wave (CW)} tone at 22 dBm into a 10 dBi horn with 55$^\circ$ half-power beamwidth in both elevation and azimuth. The receiver has a 10$^\circ$ (24 dBi) horn mounted on a rotating platform allowing a full angular scan every 200 ms with 1$^\circ$ angular sampling. The receiver records power samples at a rate of 740 samples/sec, with a 20 kHz receive bandwidth and effective noise figure of 5 dB. The system was calibrated to assure absolute power accuracy of 0.15 dBm. The high dynamic range of the sounder allows reliable measurement of path loss up to 171 dB with directional antenna gains. A detailed description of the sounder can be found in \cite{WCL_13}.
	
 \subsection{Measurement environment}\label{sec:scenario}

Measurements were performed in a New Jersey suburban residential neighborhood of mostly 2-story single-family homes along streets about 25 meters wide. The vegetation along the streets and on the properties consisted of tall trees and bushes, as representative of North-Eastern US, with single-family homes typically constructed using softwood lumber for framing, plywood or strand board for exterior walls, and drywall for interior walls. Such materials suffer considerably less penetration loss as compared to brick/concrete walls \cite{WCL_10}. 

\begin{figure} 
	\centering
		\includegraphics[width=0.85\figwidth]{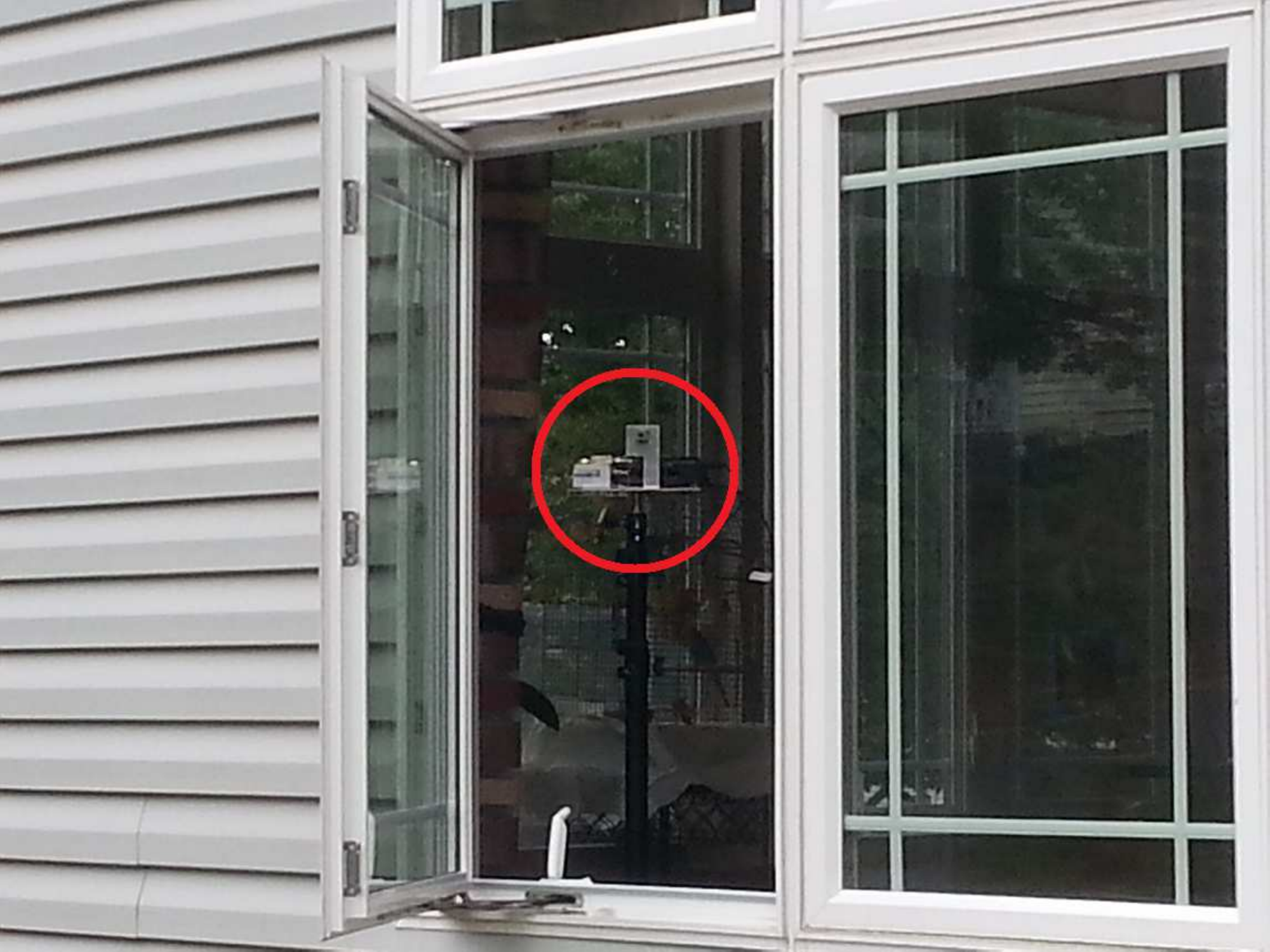}\\
		\includegraphics[width=0.85\figwidth]{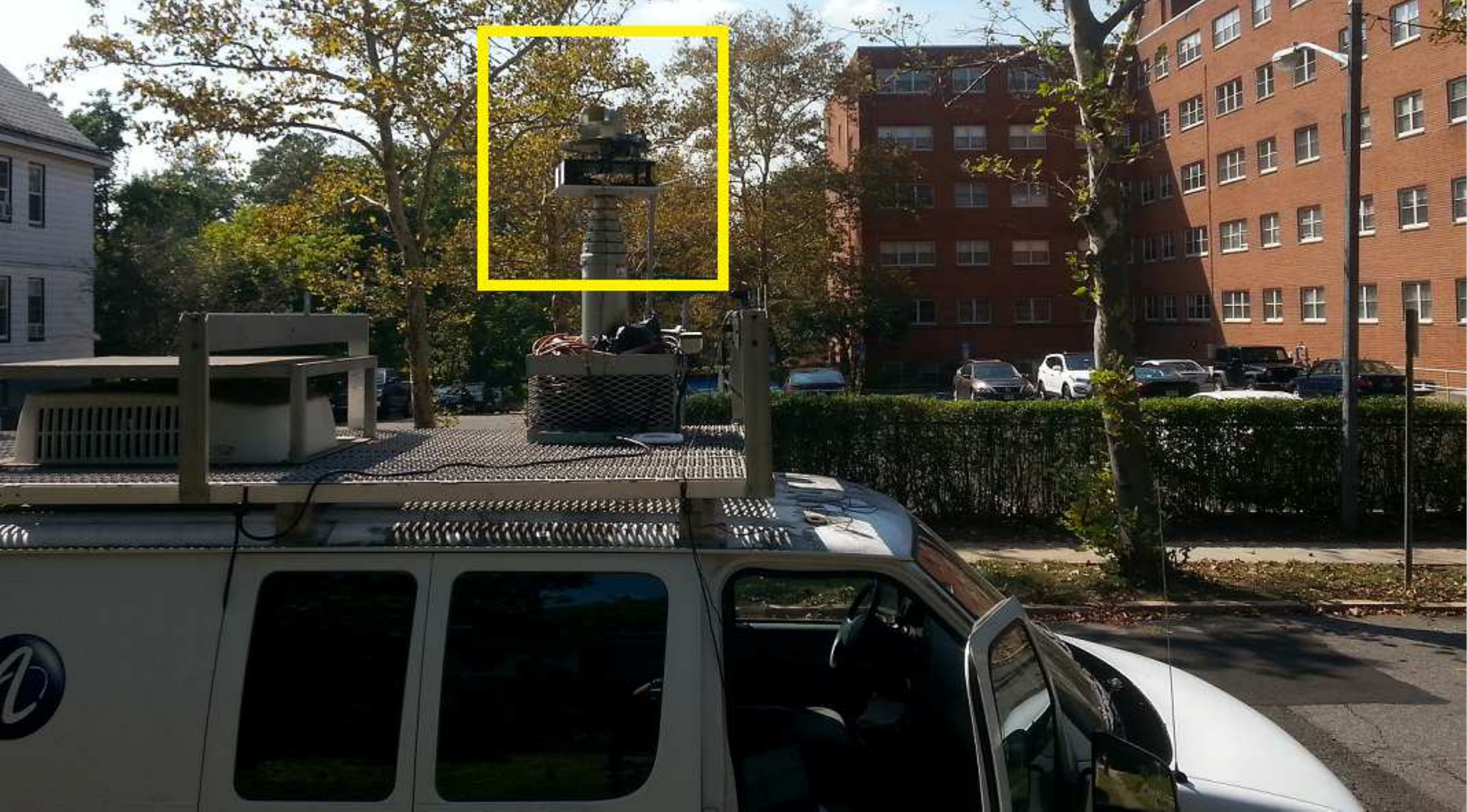}
	\caption{The fixed 55$^\circ$ horn antenna (upper) was mounted either on the exterior of a street-facing window or inside, 1.5 meters behind the closed window at the same height. The rotating 10$^\circ$ horn (lower) was on the top of a van mast, 3 meters above ground, emulating a base station on lamppost.}
	\label{fig:equipment}
\end{figure}

\begin{figure} 
	\centering
		\includegraphics[width=\figwidth]{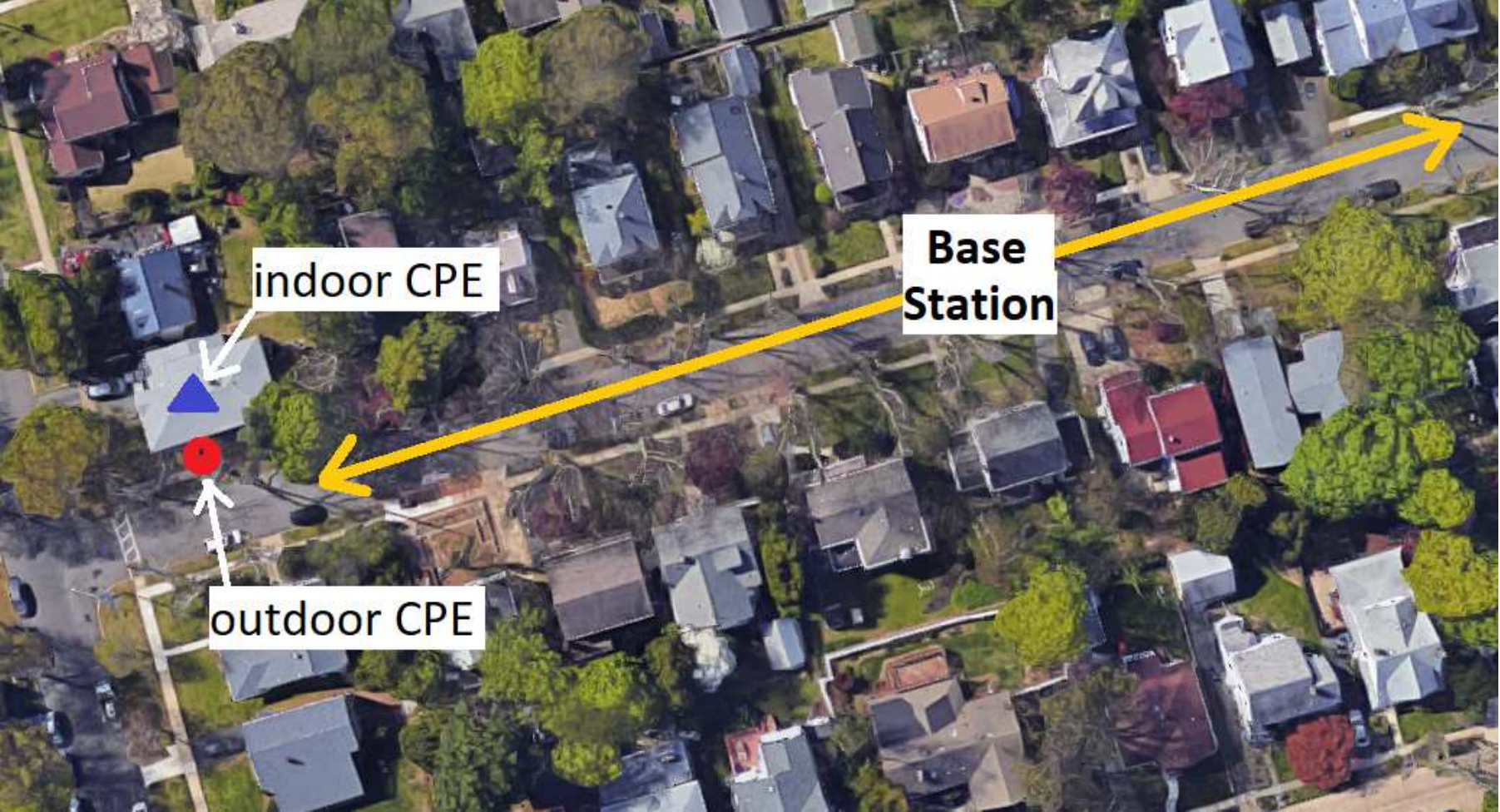}
	\caption{A street measurement map in NJ suburb. \textcolor{\editcolor}{The outdoor CPE (indicated by the red circle) was aimed 45$^\circ$ to illuminate the street, and the indoor CPE (blue triangle) was aimed perpendicular to the window}. The rotating horn ``base station''  moved along the street at ranges from 20 to 200 meters (orange line trajectory).}
	\label{fig:street}
\end{figure}

As shown in Fig.~\ref{fig:equipment}, the receiver used a spinning horn antenna, mounted on the top of a van mast 3 meters above ground to emulate a base station on a lamppost. It was driven along the street, stopping every 3 meters to collect measurements at ranges from 20 to 200 meters\footnote{Due to logistic constraints, measurement at the first house was up to a range of 100 meters in 1-meter steps.}, \textcolor{\editcolor}{where direct paths to the home} were blocked by trees and bushes. Each link measurement lasted 10 seconds consisting of over 37 full azimuthal scans, each with power recorded as a function of time and  azimuth angle \textcolor{\editcolor}{to allow small-scale fade mitigation~\cite{WCL_13}}.
 The CPE used a Tx horn mounted either on the exterior of a street-facing window \textcolor{\editcolor}{aiming at 45$^\circ$  (bisecting the 90$^\circ$ angle between the normal to the house wall and the direction along the street) to illuminate the street towards the base station~\cite{WCL_15},} or at the same height 1.5 meters inside, \textcolor{\editcolor}{aiming perpendicular to the window, which was generally found to minimize  penetration loss}. A typical measurement geometry is illustrated in Fig.~\ref{fig:street}.  
\textcolor{\editcolor}{Measurement to characterize indoor propagation and angular spread caused by nearby objects is interesting, and we leave that for future work.}

 \subsection{Data processing}

For each measured link, path gain was computed based on averaging power over all angular directions and over time (more than 37 complete azimuth scans). Since angular average provides local average power as would be obtained from a spatial average of omni antenna measurements \cite{WCL_13}, the power-average operation at each link mitigates small-scale fading and averages out power fluctuation caused by wind-blown leaves. 
	Location-dependent power variation should also be mitigated to account for different possible locations of base station placement along the street. At every measured home we have two datasets, one for indoor CPE and one for outdoor, each containing 50 or more measured links distributed evenly along the street.  \textcolor{\editcolor}{Since the street-dependent outdoor path loss  is} well represented by the power-law model with log-normal variation \cite{WCL_14}\cite{WCL_15},  \textcolor{\editcolor}{and  similar slopes (close to 4) were obtained by slope-intercept fitting to indoor and outdoor datasets, which in general exhibited a peak along the street direction,} we propose a \textit{common-slope cross-comparison} approach to separate the building penetration loss from \textcolor{\editcolor}{outdoor path loss.} 
	
	Let $P_{in}$  and $P_{out}$  be the measured path gain for indoor and outdoor terminal placement, respectively, we apply the slope-intercept fit to both datasets to determine a common slope for that street and a unique intercept for each dataset. The average path gain at distance $d$ meters is modeled as
\begin{equation}
\begin{aligned}
f_{in}(d) & = b_{in} + n*10*\log_{10}(d) \mbox{ [dB]}, \\
f_{out}(d) & = b_{out} + n*10*\log_{10}(d) \mbox{ [dB]},
\end{aligned}\label{eqn:fit}
\end{equation}
where $(n, b_{in}, b_{out})$  are chosen to minimize the fitting error
\begin{align*}
\delta = \sum_i(P_{in}(d_i){-}f_{in}(d_i))^2 + \sum_j (P_{out}(d_j){-}f_{out}(d_j))^2.
\end{align*} 
%
Each measured path gain for indoor (resp. outdoor) terminal placement was then compared against the outdoor (resp. indoor) slope-intercept fit in \eqref{eqn:fit}, and the gap was recorded as one sample of building penetration loss (BPL):
\begin{align}
\mbox{BPL} =\{f_{out}(d_i)-P_{in}(d_i), \  P_{out}(d_j)-f_{in}(d_j)|\forall i, j\}. 
\end{align} 
We  computed the cumulative distribution function (CDF) of the ensemble of BPL samples and determined its median and mean. It is straightforward to show that the mean equals the gap between the two fitted intercepts $b_{in}$ and $b_{out}$. The proposed approach avoids large variation in BPL samples that would have been observed by point-wise comparison, and it introduces negligible\footnote{\textcolor{\editcolor}{In all the three measurement trials reported here, the common-slope regression for both indoors and outdoors introduced less than 0.1 dB increase in RMS fitting error.}} increase in RMS fitting error as compared to optimizing both slopes separately.

\section{Measurement results}\label{sec:results}

\begin{figure} 
	\centering
		\includegraphics[width=0.98\figwidth]{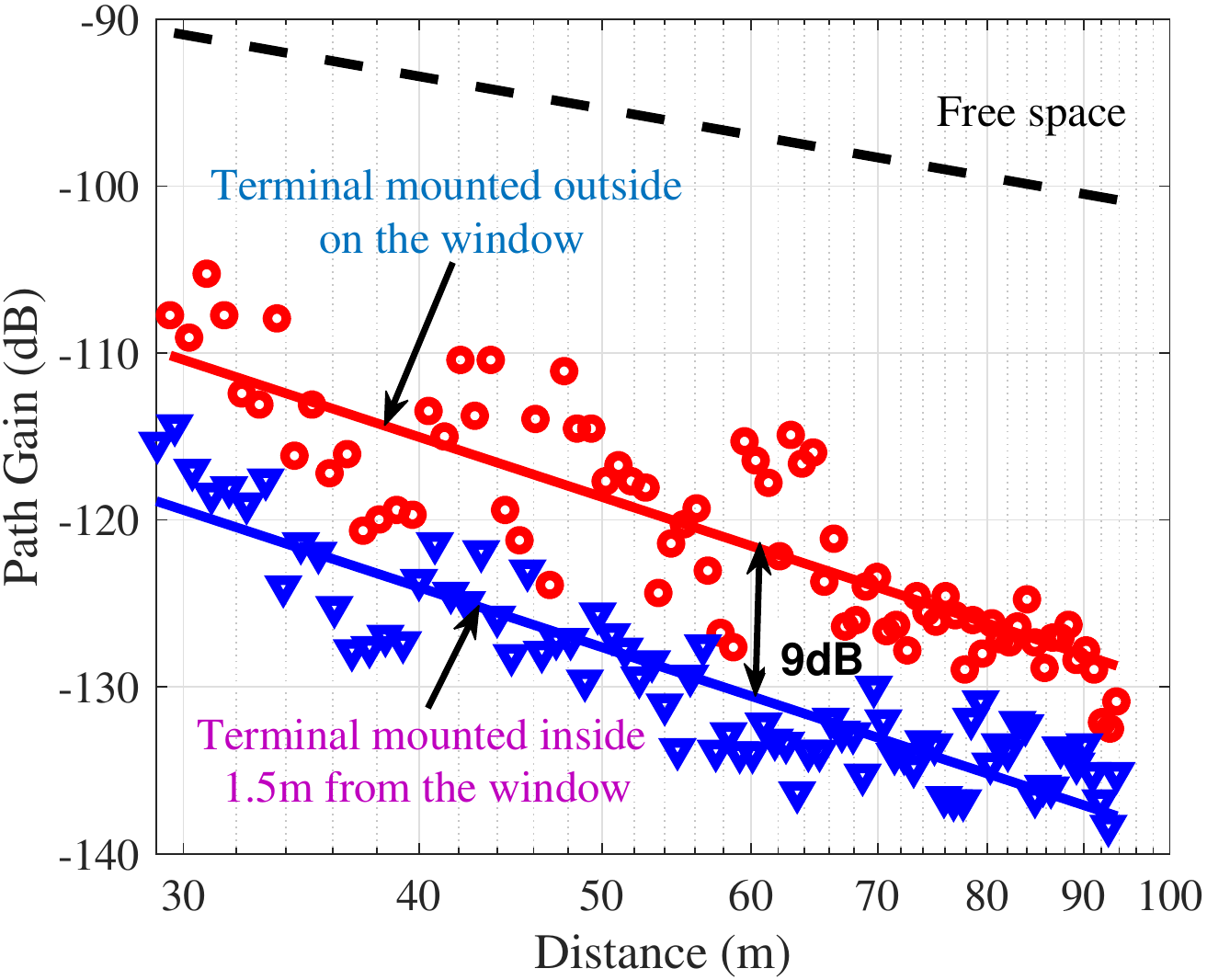}
	\caption{Measured path gain of window-mounted CPE (red circles) and indoor CPE (1.5 meters behind the window, blue triangles) for a single-family home with low-loss building materials and plain-glass windows. The average building penetration loss of 9 dB is the gap between the two fitted lines with a common slope  of 3.7 and joint RMS variation 3.0 dB.}
	\label{fig:low-loss}
\end{figure}

In Fig.~\ref{fig:low-loss} we plot the measured path gain for a house with conventional low-loss building materials and windows using two-layer plain glass. The metal mesh insect screen was removed from the window when conducting the indoor measurement. The base moved meter-by-meter down the street up to 100 meters. The slope-intercept fit provides a common slope of 3.7 and RMS fitting error 3.0 dB, \textcolor{\editcolor}{as compared to a RMS error of 2.94 dB when optimizing both slopes separately}. The gap  between indoor and outdoor intercepts is 9 dB, \textcolor{\editcolor}{which is  comparable to the  10.6 dB average loss reported in~\cite{WCL_7} where indoor terminals were placed up to 3 meters behind windows}.

\begin{figure} 
	\centering
		\includegraphics[width=0.98\figwidth]{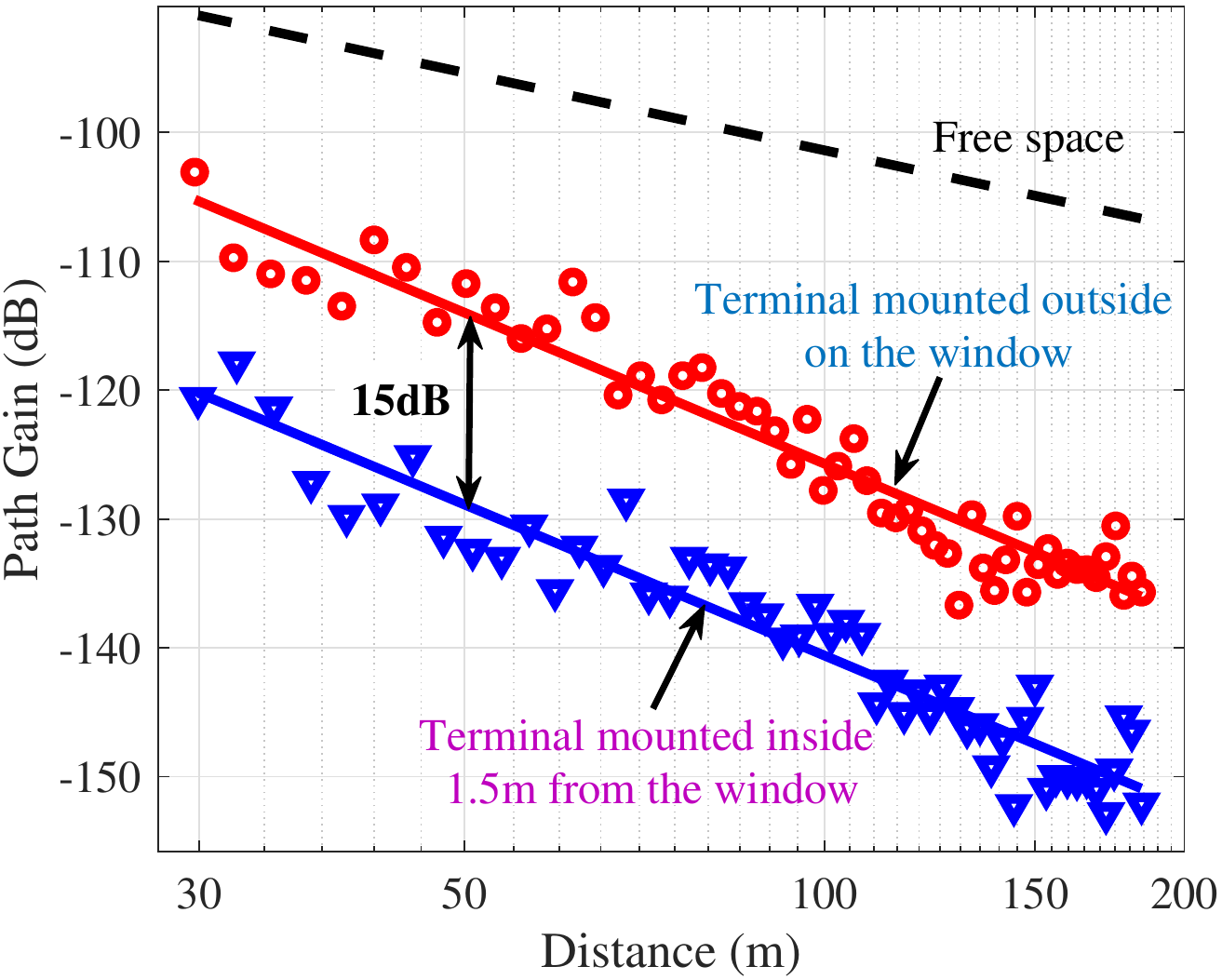}
	\caption{Measured path gain for a renewed single-family home with low-loss building materials and low-emissivity windows. The average building penetration loss was found to be 15 dB, with a common slope of 3.9 and joint RMS variation 2.5 dB.}
	\label{fig:low-E}
\end{figure}

	In Fig.~\ref{fig:low-E} we present results for a renovated  house where  low-emissivity windows had been adopted to increase energy efficiency. The base moved in 3-meter steps down the street up to a range of 200 meters. The average building penetration loss was found to be 15 dB, and the fitted lines have a common slope of 3.9 and joint RMS variation 2.5 dB.
	
\begin{figure} 
	\centering
		\includegraphics[width=0.98\figwidth]{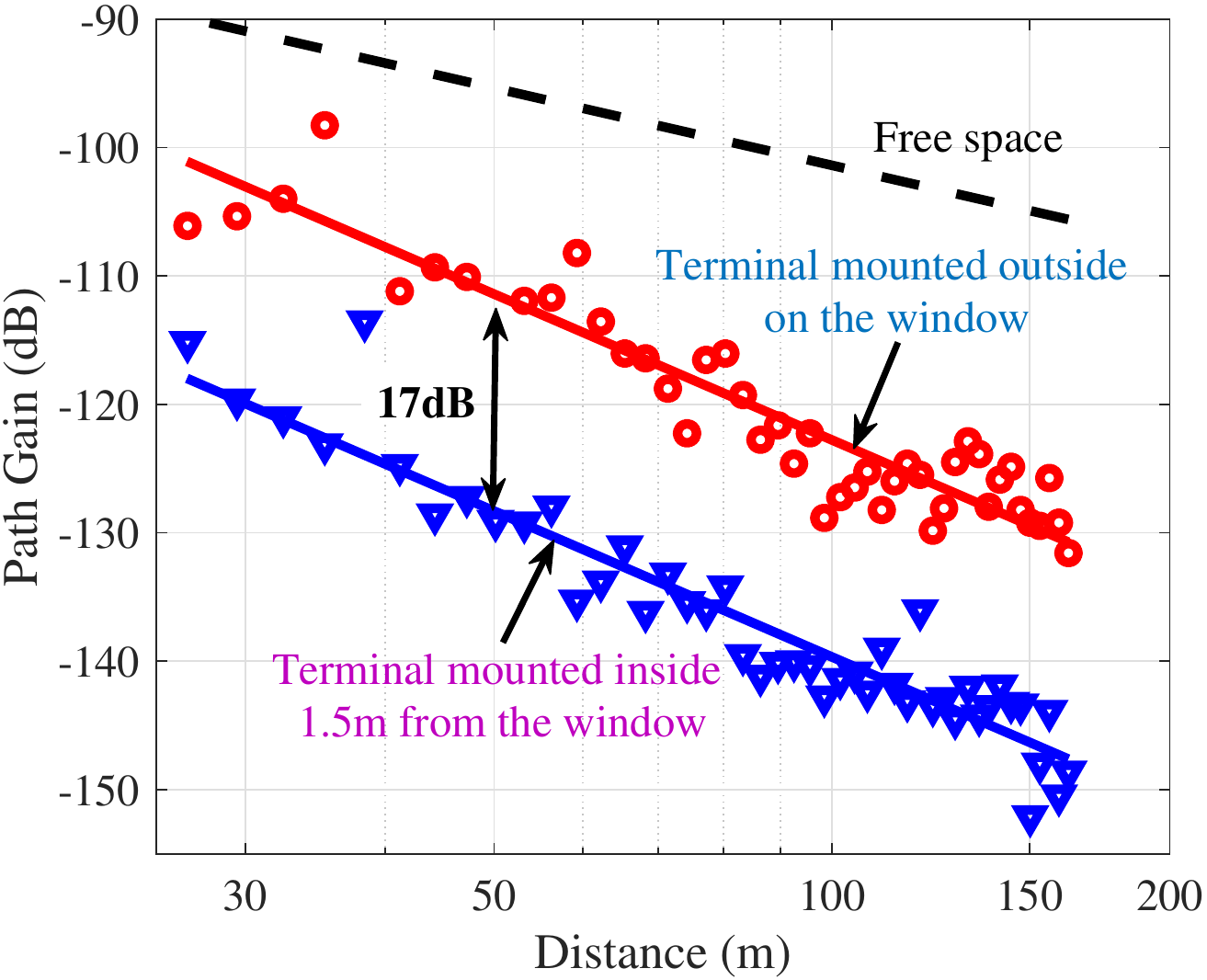}
	\caption{Measured path gain for a newly constructed single-family home with foil-backed insulation and low-emissivity windows. The average building penetration loss was found to be 17 dB, with a common slope of 3.8 and joint RMS variation of 2.8 dB.}
	\label{fig:high-loss}
\end{figure}
	
	For newly constructed single-family homes, materials with better energy efficiency have been adopted. The third house we measured is representative of such type of construction, having low-emissivity windows and foil-backed insulation. The measured path gain and regression lines are shown in Fig.~\ref{fig:high-loss}. The proposed method determined a common slope of 3.8 with a joint RMS variation of 2.8 dB. The average building penetration loss was found to be 17 dB.

\begin{figure} 
	\centering
		\includegraphics[width=\figwidth]{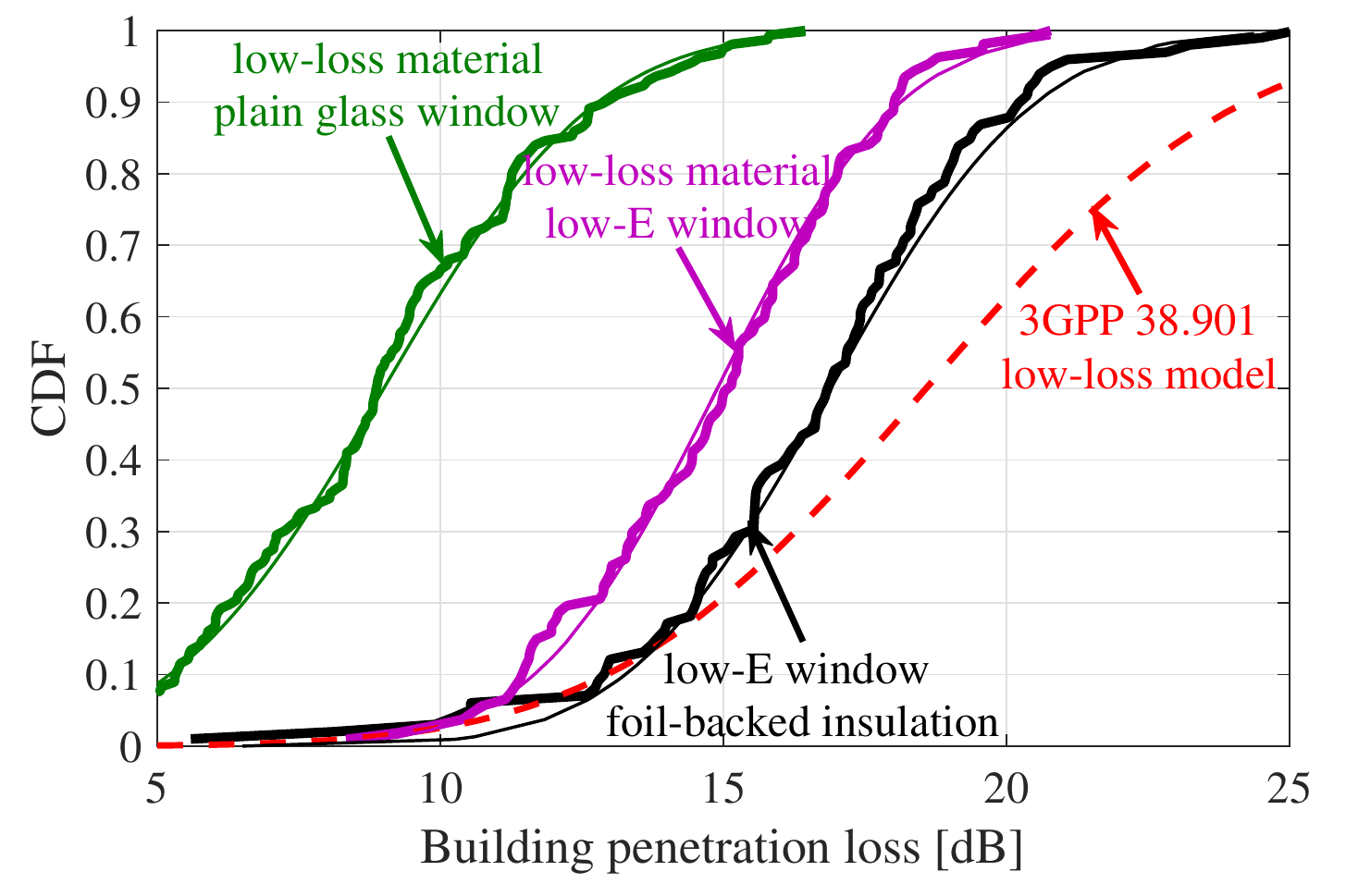}
	\caption{CDFs of measured building penetration loss for three representative single-family homes. The median was found to be 8.9 dB, 15.1 dB and 17.1 dB, respectively, as compared to the 18.6 dB median of the 3GPP low loss model. The difference between the median and mean values is within 0.2 dB, in line with the 0.15 dB power resolution of the sounder. \textcolor{\editcolor}{In all three cases the measured samples  are well represented by log-normal distributions (thin solid lines) with standard deviation of 3.0, 2.5, and 2.8 dB, respectively.} }
	\label{fig:CDFs}
\end{figure}

	In Fig.~\ref{fig:CDFs} we compare the CDFs of building penetration loss for the three scenarios, where the median building penetration loss, quantified based on the CDFs of path gain difference, was found to be 8.9 dB, 15.1 dB and 17.1 dB, respectively, which are within 0.2 dB from their corresponding mean values.  \textcolor{\editcolor}{Measured building penetration loss in all three cases can be well represented by log-normal distributions.} 
	\textcolor{\editcolor}{Note that 3GPP also adopts log-normal distributions for building penetration loss, but its low-loss model\cite{WCL_11}, which corresponds to a building with material composition of 70\% concrete and 30\% standard multi-pane glass}, has a median loss of 18.6 dB. For all the three measurement scenarios reported here, wood is the dominant building material which has significantly less penetration loss than concrete. \textcolor{\editcolor}{Therefore, the 3GPP low-loss model can be refined to reduce overestimation of penetration loss for specific suburban scenarios.}
	

\section{Conclusions}\label{sec:conclusion}

 We reported measured building penetration losses in a NJ suburban residential neighborhood for fixed wireless access deployment, where FWA base stations were mounted at a height of 3 meters to mimic lamppost placement along streets. The terminal emulating CPEs was positioned either on the exterior of a street-facing window or 1.5 meters behind it inside each house, acting as relay to provide indoor coverage throughout the desired area. Small-scale fading and time fluctuation were averaged out by power average at each measurement location over angle and time, and large-scale variation was averaged out by the proposed \textit{common-slope cross-comparison} method. Three representative single-family homes were measured, and the median building penetration loss was found to be 9 dB for a home with low-loss materials and plain-glass windows, 15 dB for a renovated home with low-emissivity windows, and 17 dB for a new construction using foil backed insulation and low-emissivity windows. 

\textcolor{\editcolor}{The reported 9 to 17 dB  building penetration loss can be interpreted as the  additional loss for outdoor signals to reach the indoor CPE that is positioned 1.5 meters behind the window facing the base station. Much higher penetration losses are to be expected when trying to cover most of the indoor space from outdoors. Since the outdoor path loss at 28 GHz in suburban residential areas is already challenging~\cite{WCL_15} owing to vegetation blockage and under-clutter propagation, FWA to indoor CPE is more likely for low-loss residential homes and for homes that are at a short distance to lamppost-mounted base stations.}

\end{document}